# Introducing Automated Regression Testing in Open Source Projects

Christopher Oezbek
Freie Universität Berlin
Institut für Informatik
Takustr. 9, 14195 Berlin, Germany
christopher.oezbek@fu-berlin.de

## Abstract

*To learn how to introduce automated regression testing to existing medium scale Open Source projects, a long-term field experiment was performed with the Open Source project FreeCol. Results indicate that (1) introducing testing is both beneficial for the project and feasible for an outside innovator, (2) testing can enhance communication between developers, (3) signaling is important for engaging the project participants to fill a newly vacant position left by a withdrawal of the innovator. Five prescriptive strategies are extracted for the innovator and two conjectures offered about the ability of an Open Source project to learn about innovations.*

## Contents

## 1 Introduction

The Open Source development paradigm based on copyleft licenses, global distributed development and volunteer participation has become an alternative development model for software, competing on par with proprietary solutions in many areas. Open Source software especially has established a good track record related to quality measures such as number of post-release defects or time to resolution for bug reports [50, 57]. These achievements are typically seen as hailing from open access to source code and from openness to participation, which leads to increased peer review and involvement of bug-fixing experts as summarized in Raymond's quote "Given enough eyeballs, all bugs become shallow" [60].

To further improve the quality assurance methods employed in Open Source projects and learn about the mechanisms by which Open Source projects adopt new innovations in general, the introduction of automated regression testing into an Open Source project is studied. Regression testing was chosen because it represents a well-known and established quality assurance practice from industry. The perspective assumed in this study is from an individual on the outskirts of the project who has used the project but not actively participated and who aims to strengthen the quality attributes of the project. This could be a developer in a company who has decided to use an Open Source software as an off-the-shelf component for building a software application and is now interested in long-term quality improvements in the project.

For this study, *introduction* is defined as the set of activities meant to achieve adoption of a technology, practice or tool with the boundaries of a social system such as a software development project. In the case of this study, the innovation is thus the practice of using automated regression testing. Adoption for this matter has been achieved, when the uses of an innovation have become "embodied recurrent actions taken without conscious thought" [13], i.e. the writing and executing of test cases in the case of regression testing have become a habit or routine.

To develop an understanding of how to introduce automated regression testing to Open Source projects and whether usage is as expected, an exploratory field experiment was conducted (the methodology is explained in full detail in Section 3). The study was performed with the Open Source project FreeCol described in more detail in Section 3.2.1 and had the following research questions in mind:

1. Is the introduction of a quality assurance process innovation such as automated regression testing a feasible possibility for an external innovator?

2. Which aspects of the introduction and adoption of an innovation need particular assistance?

3. How should the innovator behave and which strategic advice can be deduced?

4. What can be learned about automated regression testing within the scope of Open Source software development?

The remainder of the article presents a more detailed overview of the practice of automated regression testing, the project chosen to introduce testing with and the methodology used in the study. The last two sections then describe the actual introduction effort and the results achieved.

## 2 Automated testing

Automated testing is the practice of writing executable specification against which an application can automatically be tested. "Automated testing" is used as a term to distinguish from manual testing in which software response is verified by a human against manually entered input.[1] We typically distinguish unit testing of individual classes and modules, integration testing of the interaction of several modules, and system testing of the application in the context of later use [76, p.77]. Since the boundaries between these different testing scopes are often hard to draw precisely, the term unit testing, which was originally reserved for testing the smallest possible unit in isolation and allowing also manual execution, has become near synonymous with automated testing independent of the granularity of the tests contained in a test suite.

While automated testing is primarily a quality assurance technique to achieve compliance with specification, it also has become popular to view it (a) as a specification in its own right, for instance when translating an ambiguous bug report into a specific test case, and (b) as a means to track development progress by the individual developer ("you will be done developing when the test runs" [5]).

It is not necessarily defined at which time the test cases are created with relation to the code they are testing. In a waterfall or V model software development process it is likely that unit and integration tests will be created bottom-up in a "sequential fashion" after implementation and integration respectively [38] and in a dedicated testing department. Yet, with the advent of more agile development processes, developers have become increasingly involved in writing tests during or even before implementation. Such *test-driven development* (TDD, also "test-first") has received a lot of attention in academia and industry [36] and is characterized by repeated cycles of writing tests, implementing functionality to pass the tests and subsequent refactoring to improve the quality of the code.

The last step in the TDD cycle is aided by the *regression* detecting capabilities of a test suite. Since tests ensure compliance to the behavior coded in them, deviations from this behavior will cause affected tests to fail. This capability of tests is cited to increase developers' confidence to refactoring code [49, p.129].

Lastly, for the relationship between testing and finding bugs in software, the black-box properties of the test execution must be regarded. This is famously summarized by Dijktra's remark that "Program testing can be used to show the presence of bugs, but never to show their absence" [14], which reminds us that tests compare expected and actual output of the program for a defined set of scenarios and do not prove correctness of the code. In practical terms this means that even though the tester derives confidence about the quality of the code by successfully testing a program, it still might be that the tests (1) do not *cover* all the scenarios in which the code might be used, (2) contain bugs themselves, and (3) do not detect interrelating defects that result in seemingly correct behavior.

At this point we do not further qualify what kinds of means and ends are targeted by the habit of automated testing (for instance whether developers translate bug reports to test cases or use it for test-driven development), but rather leave the exact kind of usage up to the Open Source project and its specific requirements.

## 3 Methodology

This study is a long-term field experiment [31] with post-hoc data analysis being performed both quantitatively and qualitatively. The study proceeded in four steps: First some time was spent on building a theoretical model of how an innovation introduction with regard to automated testing should proceed. Second, a project was selected to conduct an innovation introduction following this model with, the result of which was the project FreeCol. Third, testing was introduced to this project, which took place from April to September 2007. Last, the outcome of the introduction was analyzed by means of data mining the source code repository of the project and analyzing the mailing-list communication on testing up until August 2009.

[1] From now on, we will use the unqualified term of *testing* as referring to *automated* testing exclusively and will qualify all occurrences of testing as pertaining to manual testing respectively.

Even though Open Source projects are robust against negative external influence, it was attempted to minimize risk toward the project and to protect the autonomy of the subjects [8, 53] by creating an atmosphere of collaboration, involvement and participation between project and researcher, and protecting privacy and confidentiality [4, 22].

### 3.1 A model of external innovation introduction

To give the innovator a plan for introducing automated testing according to which to operate, first a prescriptive stage model of introduction behavior was designed. The goal was to guide the innovator and make the process reproducible in other projects and context. The resulting model is presented in the following paragraphs and summarized in Figure 1.

The goal of the first phase is to get to know the Open Source project and establish the technical requirements of participating in the project. This entails subscribing to and reading the mailing-list of the project (*lurking* [52, 58]) with a focus on community and power structures of the project, downloading the source code from the repository, setting up the development environment, reading mission statements and task trackers, and building the project.

In the next phase the innovator is to become *active* by writing test cases and contributing them unsolicitedly to the project using the mailing-list (see [63] for a discussion on contact strategies). This strategy is to be used to create something valuable for the project [16, 72] and as a side-effect to become known and possibly gain write access to the project repository. Areas to test will be selected to maximize the learning experience for the innovator regarding the code base, while still providing benefit to the project. Previous work had shown that to be beneficial, close integration with existing code structure and coding conventions is important [59]. To make these test contributions usable for the project, documentation on testing will be written and the build-structure of the project improved to better accommodate testing. An upper limit of 10-15 hours per week in time spent on communicating and writing test code will be set, so that the researcher is not overtaxing the capacity of the project and is behaving in line with the average time commitment of OSS project participants [26].

In the third phase the focus will be shifted away from *solitary* activity, which is communicated primarily via the project leadership, to *collaboration* with other project members, based on three different collaboration models for automated testing: (1) The innovation will codify recent bug tracker entries into failing test cases and contact the developers who had previously worked in that area. (2) The innovator will add test cases for code recently committed by other members ex post. (3) The innovator will write failing test-first code for features that members are discussing to implement in the very near future. Each of these areas should generate a possibility for interaction and collaboration between one or several developers and the innovator. The goal here is to build a social network between the introducer and the developers, to spread knowledge about automated testing and demonstrate benefit of the innovation to the developers.

In the fourth phase the activity of the developer will be *phased-out* slowly with an emphasis on support and maintenance. The introducer will actively monitor test contribution by other developers and guide them to develop high quality tests with large benefit for the project.

### 3.2 Choosing a project

Using the project hosting site SourceForge.net, I first randomly chose among projects fitting the following set of criteria: (1) the project uses an Open Source license and development model (in particular this means distributed development and openness to outside contributions [7]), (2) the project has between 5 and 15 active developers who have committed code to the CVS repository within the last three months to ensure both interesting interactions and sufficient possibilities for the researcher to act upon, (3) the project uses C/C++, Java, C#, Ruby, Perl or Python, (4) the code-base contains less than 15 test cases prior to being approached, (5) the project is not in process of adopting any time-consuming innovations, and (6) the project consists not primarily of GUI code (which is harder to test [48]).

These criteria were used to ensure that the case is (a) typical enough to generalize to other projects, (b) suitable for automated testing, (c) has potential for interesting interaction regarding the introduction, and (d) automated testing is relatively new with regard to the project [54].

Ten projects were then randomly chosen from a list of projects with at least 5 developers as given by SourceForge.net and analyzed for the remaining criteria. It turned out that none of the these projects fulfilled all criteria and it was decided not to continue by randomly selecting projects (three projects were inactive in the last 3 months, one project already had more than 15 test cases, three projects were moving to another source code management system, which was seen as major innovation introduction underway, two projects did not have any public communication, and one project did have more than 5 members but consisted of several completely independent sub-projects none of which passed the five-member-hurdle). At this time, the SourceForge.net newsletter arrived which features a “Project of the Month”[2] chosen by the SourceForge.net staff. As the first project listed in this newsletter already complied with all set criteria, choosing from the list of Projects of the Month offered a suitable alternative to a random sample.

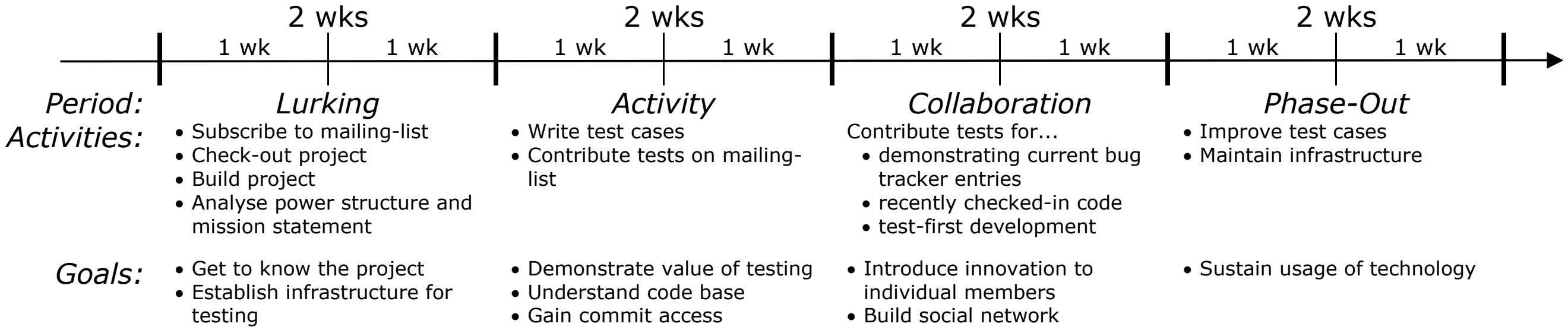

Figure 1. Phases in the introduction process.

### 3.2.1 FreeCol

FreeCol is a project striving to recreate the turn-based strategy game Sid Meier’s Colonization – a sequel to Sid Meier’s successful empire building game Civilization [23]. The project was started in March 2002 and had its first release on January 2, 2003[3]. FreeCol is structured as client-server application to facilitate multi-user play and is written entirely in Java. The project is regularly ranked in the top 50 of Open Source projects hosted at the project hoster SourceForge.net and 16,500 copies have been downloaded per month on average over the last six months. The project is lead by the two maintainers Stian Grenborgen and Michael Burschik and has 60 members enlisted on the SourceForge.net project page[4] of which 46 are designated as developers and 13 of which are deemed active[5]. The project already had one (1) test case using JUnit which was integrated into the Ant build-scripts.

## 3.3 Conducting the innovation introduction

The introduction of automated testing into the project FreeCol according to the conceptual model described in the Section 3.1 was begun at the beginning of April 2007 and lasted for six weeks until May 15, 2007. The lurking in particular was shortened to just 2 days, because I had become sufficiently familiar with the code to find a defect, provide a test case for it and fix the associated issue. I then thus proceeded directly to actively contribute on the mailing-list and was granted commit privileges within the first week of doing so. This phase of writing and contributing test code naturally evolved into the more collaborative phase 3 of my engagement. First, one of the test cases I had written caught a regression caused by a previous check-in which allowed me to start communicating with the developer who had caused the failure. Second, when new developers asked on the list about open tasks, I noted that writing test code could be a good way to get started, which convinced one new developer to provide two test cases. Third, I created several tests to show which aspects of a reported bug had not yet been fixed. Rather than fixing the bug myself to make the tests pass, I advertised it as opportunity for working on a well specified problem. After 4 weeks of engagement the number of test cases in FreeCol had then increased from 1 to 57 (see Figure 8) and this was deemed sufficient for phase 4 to start. As part of phasing out over the next two weeks, I added more documentation regarding testing, created a short video for setting-up Eclipse to run the test suite for FreeCol and fixed the Ant build-scripts to execute the test cases for those not using Eclipse. May 15th marks the last day of the phase-out.

Returning at the beginning of June and July each, I had to notice that the introduction so far was unsuccessful with no new tests having been created and the number of tests being stuck at 65 tests. Two months later, in August 2007, one of the project maintainers informed me that the test suite was now “spectacularly broken” after a large feature addition and refactoring had changed substantial parts of the software (see [71] for an overview how refactoring and testing relate to each other). During a discussion about the usefulness of testing, I then agreed to repair the test suite, which was finished mid September 2007, before ending my engagement as a committer in the project.

Over the next two years, my engagement on the mailing-list was restricted to answering several questions with regard to using Eclipse for developing FreeCol and running the tests using Ant. When returning to the project in August 2009 and analyzing both the repository and mailing-list, it turned out that the number of test cases had increased markedly, which will be discussed in Section 4 on Results after the data analysis methodology has been presented below.

[2] http://sourceforge.net/community/potm/

[3] http://www.freecol.org/history.html

[4] http://sourceforge.net/project/memberlist.php?group_id=43225

[5] http://www.freecol.org/team-and-credits.html

### 3.4 Data analysis methodology

The ex-post data analysis of the introduction of testing was performed using two primary data sources. First the source code repository of FreeCol was mined[6] to discover the number of test cases in FreeCol, failure rates and statement coverage (see [80] for an overview regarding measuring test case coverage). Second, e-mail lists were downloaded from SourceForge.net as permitted by one of the project maintainers and were analyzed briefly and qualitatively for social interaction regarding testing.

The detailed steps of the repository mining were:[7]

1. To perform data analysis efficiently, the FreeCol source code repository (managed via Subversion [9]) was first mirrored locally using `svnsync`[8].

2. An XML version history log was next extracted for the whole repository and converted to a CSV file including revision, author, commit messages, date and affected paths via xml2csv[9]. Using a regular expression search, the commits which affected test classes were identified, the result of which was appended as a last column to the table.

3. This data was then used to produce Figures 2, 3, 4 using the R project for statistical computing [61].

4. Using a Ruby [19] script as a driver, FreeCol was then checked out with the appropriate version at each month since April 2007. For each check-out, a customized and overlaying Ant build-script[10] was run, which would compile and then execute the test cases, thereby generating JUnit test results. Test coverage was calculated using Cobertura[11]. Using the Nokogiri XML API[12], tests results were converted to CSV.

5. The concluding analysis in R resulted in Figures 5, 6, 7, 8.

[6] See [39] for a survey of mining source code repositories, [79, 81, 24, 47, 46] for selected articles, and [1] for the dangers associated with repository mining

[7] All scripts used for producing the results in this study as well as intermediate data to reproduce the statistical analysis is available at http://www.inf.fu-berlin.de/inst/ag-se/pubs/test-intro2009data.zip

[8] http://svn.collab.net/repos/svn/trunk/notes/svnsync.txt

[9] http://www.a7soft.com/xml2csv.html

[10] http://ant.apache.org

[11] http://cobertura.sourceforge.net

[12] http://nokogiri.rubyforge.org/

## 4 Results

If we first look back quantitatively at the introduction from April 2007 to August 2009, we find it a clear success: First, the number of test cases has increased almost constantly from 73 at the departure of the innovator to 277 as of August 2009. Figure 8 shows this linear growth of on average 9.9 testcases being added per month (95% confidence interval: 6.3–13.4) with no noticeable plateau. Supporting this linear growth is the percentage of monthly commits affecting test cases which is between 10.0% to 15.1% with 95% confidence, depicted in Figure 3) and implies that writing novel tests was a constantly exercised practice in the project. Second, existing test cases were maintained to execute successfully. This indicates that besides writing new test cases, the developers assumed the responsibility of keeping the test suite in working condition. Only in rare cases did the monthly snapshot copies reveal broken test cases, such as in September 2007 when all tests were broken, or in March 2008 when 15 out of 234 test cases failed (see Figure 8). Third, writing test code is not an activity performed only by one or two developers. Out of 32 developers having ever committed to the FreeCol repository for a total of 5,030 commits and thus disregarding patches submitted by other developers and committed for instance by the maintainer, 16 made at least one of the 449 test-affecting commits in their career at FreeCol (11 at least 5, 6 at least 10, 4 at least 20 commits). An overview of testing activities by developer is shown in Figure 4. Fourth, while the project grew from 31,800 non-comment source lines of code (NCSLOC) to 48,200[13] over the course of the last two years, the associated code being covered by tests was expanded from 200 to 11,000 lines (see Figure 5), which represents an increase from 0.5% coverage to 23% (see Figure 6).

Moving away from the quantitative assessment of the source code repository to an analysis of the communication on the mailing-list, we also note multiple incidents in which developers voiced their positive attitude towards testing. For example, after a successful bug resolution episode, one core developer asked whether additional test cases should be added and one of the maintainers answered enthusiastically:

> That goes without saying! We should have unit tests for absolutely everything. Everyone gets extra bonus points for writing unit tests. [freecol:2518][14]

[13] Source code lines for this study only include executable lines, i.e. this excludes documentation, whitespace and tests, but also method signature and class definitions.

[14] Citations such as [freecol:2518] are hyperlinks to the respective e-mails from the Freecol Developer Mailing-list and are numbered in the order the e-mails appeared in the mbox archive file from SourceForge.Net.

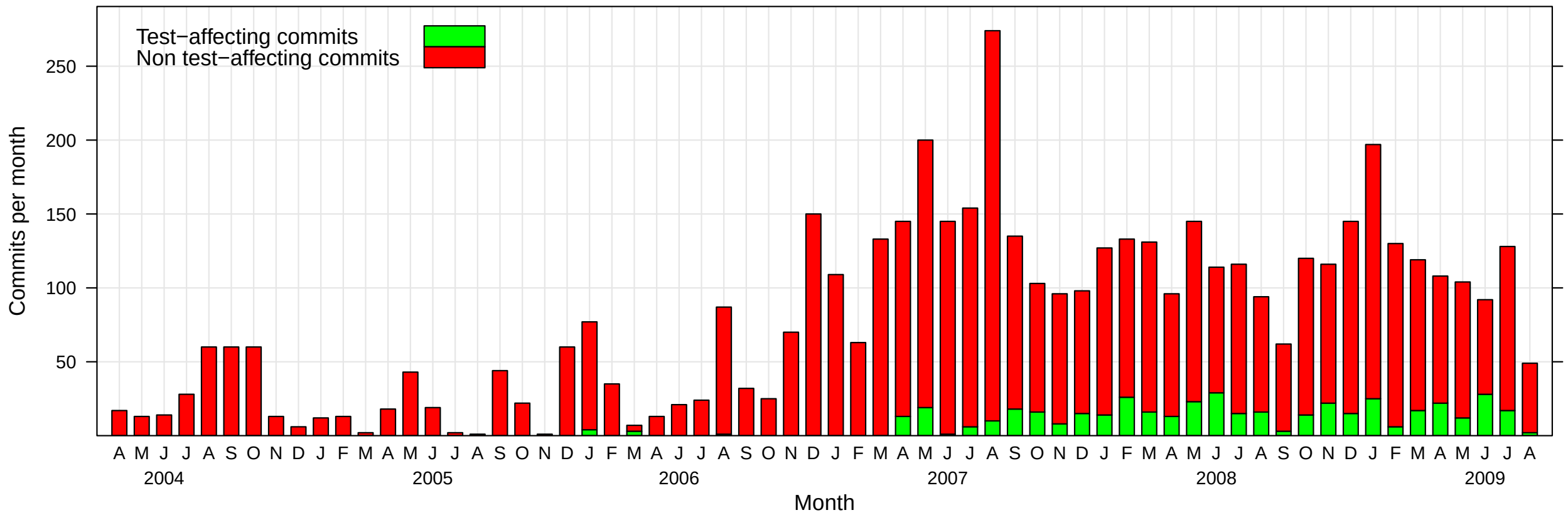

Figure 2. A time series display of the number of commits per month since the beginning of the project being managed in a source code repository. Commits which affect test cases are shown as a subset in green. It is well visible that testing activity in the project is on-going since April 2007.

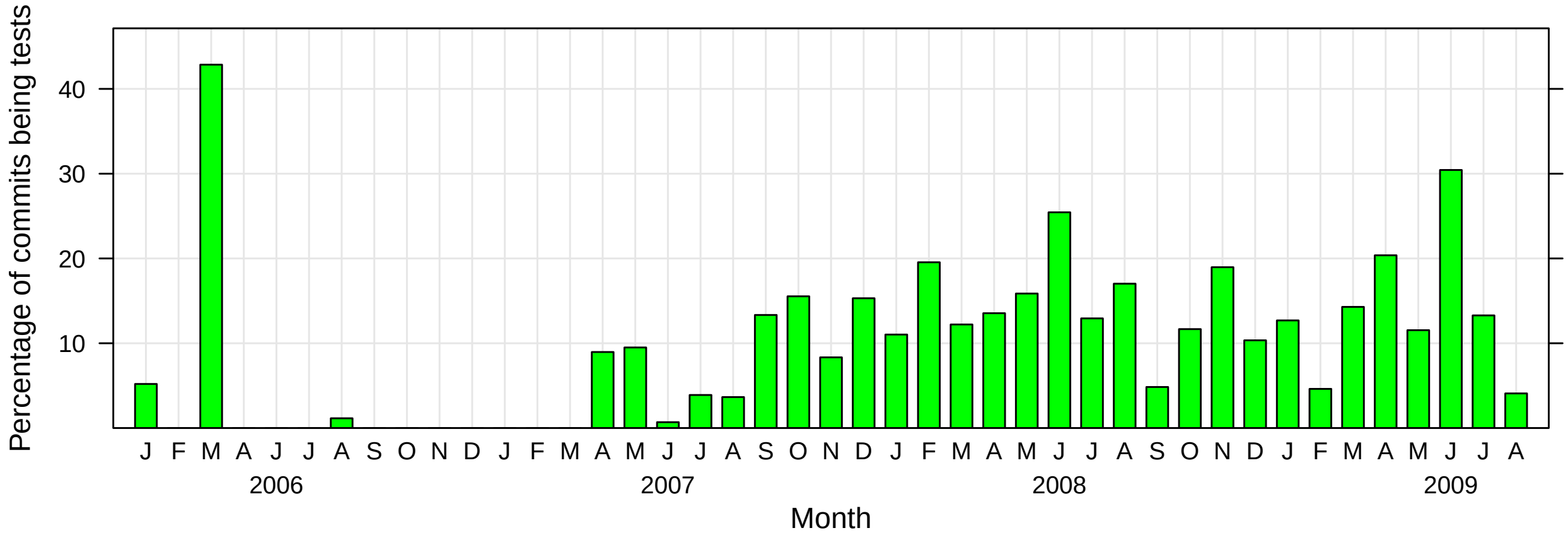

Figure 3. A time series display of the percentage of commits affecting test cases per month since the beginning of the project. It can be seen that testing activity per month since April 2007 was on average above 10% (mean 12.5%, first quartile at 8.9%, third quartile at 15.5%)

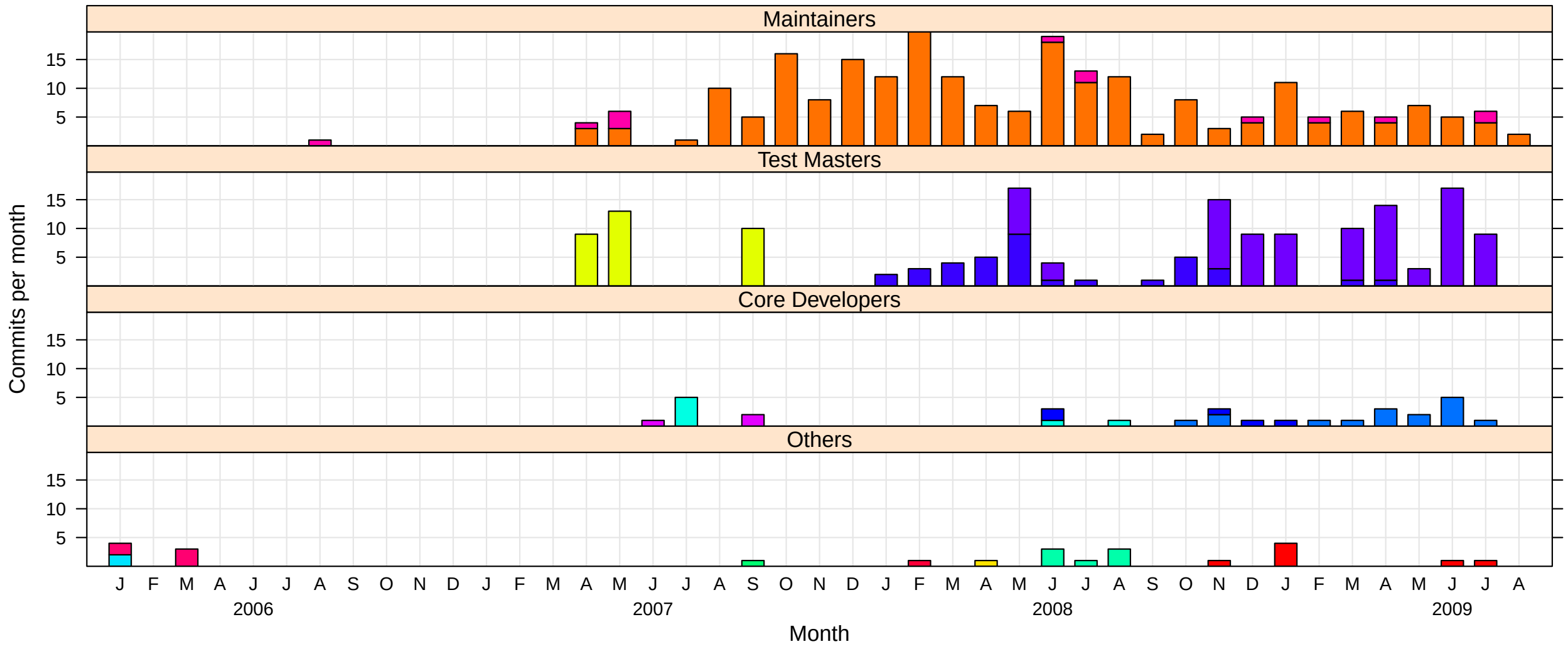

Figure 4. A time series display of absolute number of commits affecting test cases per month separated for each developer with commit access in the project. The developers have been split into the four groups of “maintainers”, “test masters”, “core developers” and “others”. A test master for this matter was defined as somebody with more than 30 test commits and a core developer as a developer with more than 80 total commits. All test masters were also core developers. The engagement of the innovator is well visible as two initial bursts of testing activity in April/May of 2007 and then in September 2007. Also the complementary patterns of activity between one of the maintainers with the test masters can be seen, when his testing activity increased during winter 2007/2008 and late summer 2008 to offset the notable absence of testing commits from the test masters. The display was cropped for the activity of one of the maintainers who made 22 test affecting commits in February 2008.

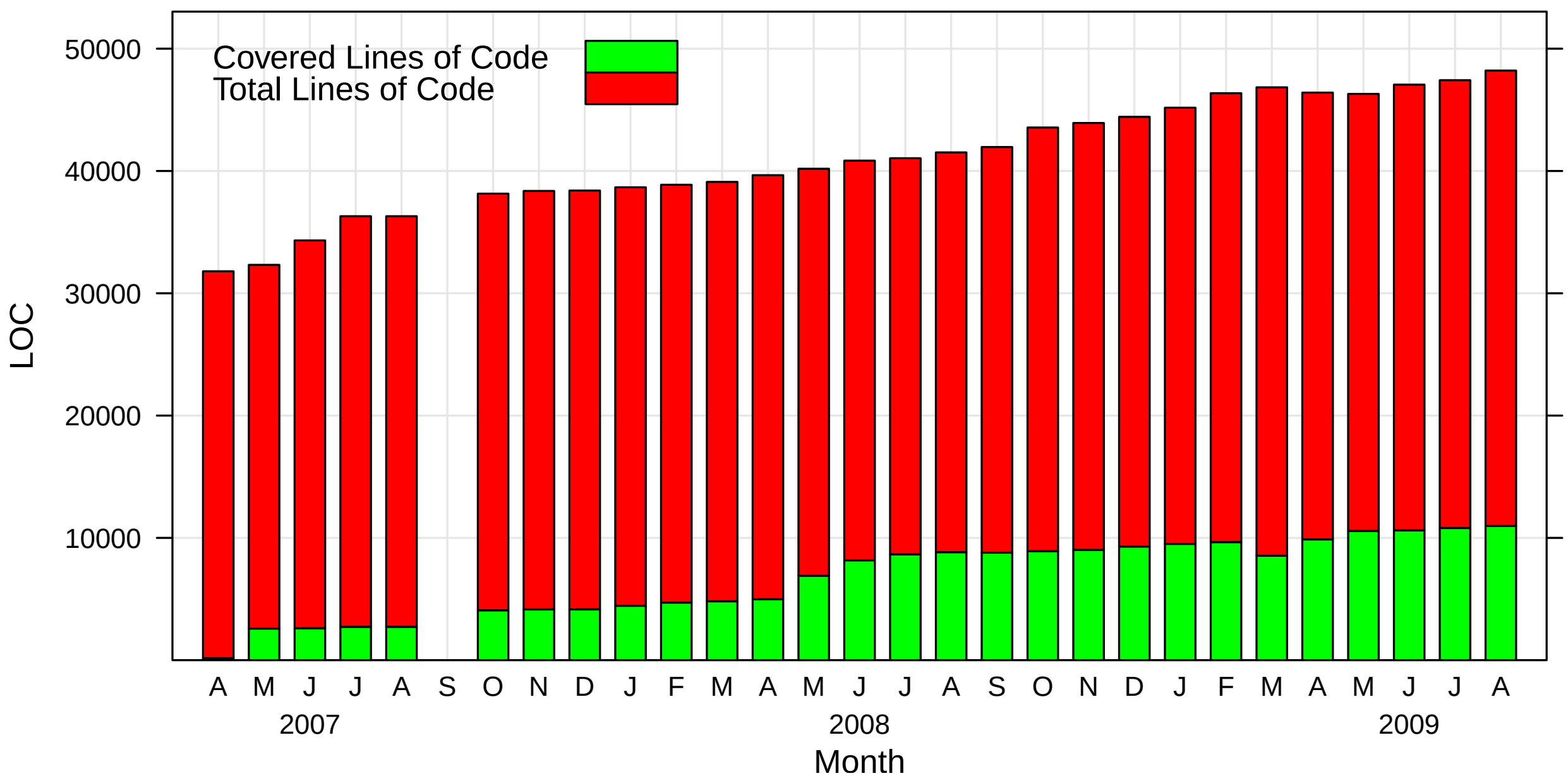

Figure 5. A time series display of the test coverage achieved over the course of the innovation introduction. Coverage is shown as the number of lines of the project's source code which were executed as part of running all tests which could be compiled for the first commit of each month displayed. We can see that while the total lines of (non-documented, non-test, non-whitespace) source code for the software increased from 31,800 lines to 48,200, the number of lines exercised by testing code increased from 200 to 11,000.

In another example, one core developer was able to catch several bugs in the existing code during the implementation of a new feature, which made him praise the test cases:

> [I] Also found a couple of existing bugs (while implementing the tests, god bless them) [freecol:3351]

While this assessment gives us reasonable certainty to declare the introduction of automated testing into the project FreeCol a success, we should now look for more general insights to be gathered from the case study. These insights can be split into (1) results regarding automated testing and (2) results regarding innovation introduction, which will be discussed in turn.

### 4.1 Insights into automated testing

The most interesting insight regarding the use of testing in Open Source projects is that test cases have been used repeatedly to enhance communication in the project in two major ways: (1) If facing a defect in the software without the necessary knowledge to repair it, or even when unable to understand the general problem, we have seen developers write failing test cases which reproduce or narrow down the failure caused by the defect and use the test case as a more concise alternative for communicating the circumstances of the failure (for instance [freecol:2606] [freecol:2610] [freecol:2640] [freecol:2696] [freecol:3983]). (2) When facing ambiguity about how FreeCol should behave, we have seen developers codify their opinion as test cases [freecol:3276] [freecol:3056] or existing tests being the starting-point for discussions about how FreeCol should behave [freecol:1935].

Thus, test cases become explicit articulations of intent during requirements arbitrage inside the project. If we abstract, we can say that in both cases the developers construct a.) an expectation against the behavior of the software but also b.) an expectation towards the behavior of the fellow project members as to fix the failing test cases. This latter aspect of communicating expectations seems to be a second major advantage beside the regression detecting abilities of having a test suite in an Open Source project (see for instance [freecol:3961] or [freecol:4431]).

It is outside the scope of this research to assess the relative advantage of communicating an informal and vague bug description or aspect of a specification using an executable test case beyond the observed episodes on the mailing-list. In particular, it would be hard to extract how information has flown inside the project and deduce the role that the test cases actually played from commit messages and e-mails on the mailing-list. For instance, while we have seen defects encoded as test cases often being fixed within 24 hours [freecol:3057] [freecol:2610] [freecol:2606], we also have at least one case in which the defect highlighted by a test was fixed while this test was just being written [freecol:2640]. Working out the mechanisms and actual pathways of causation would require quite some detective work.

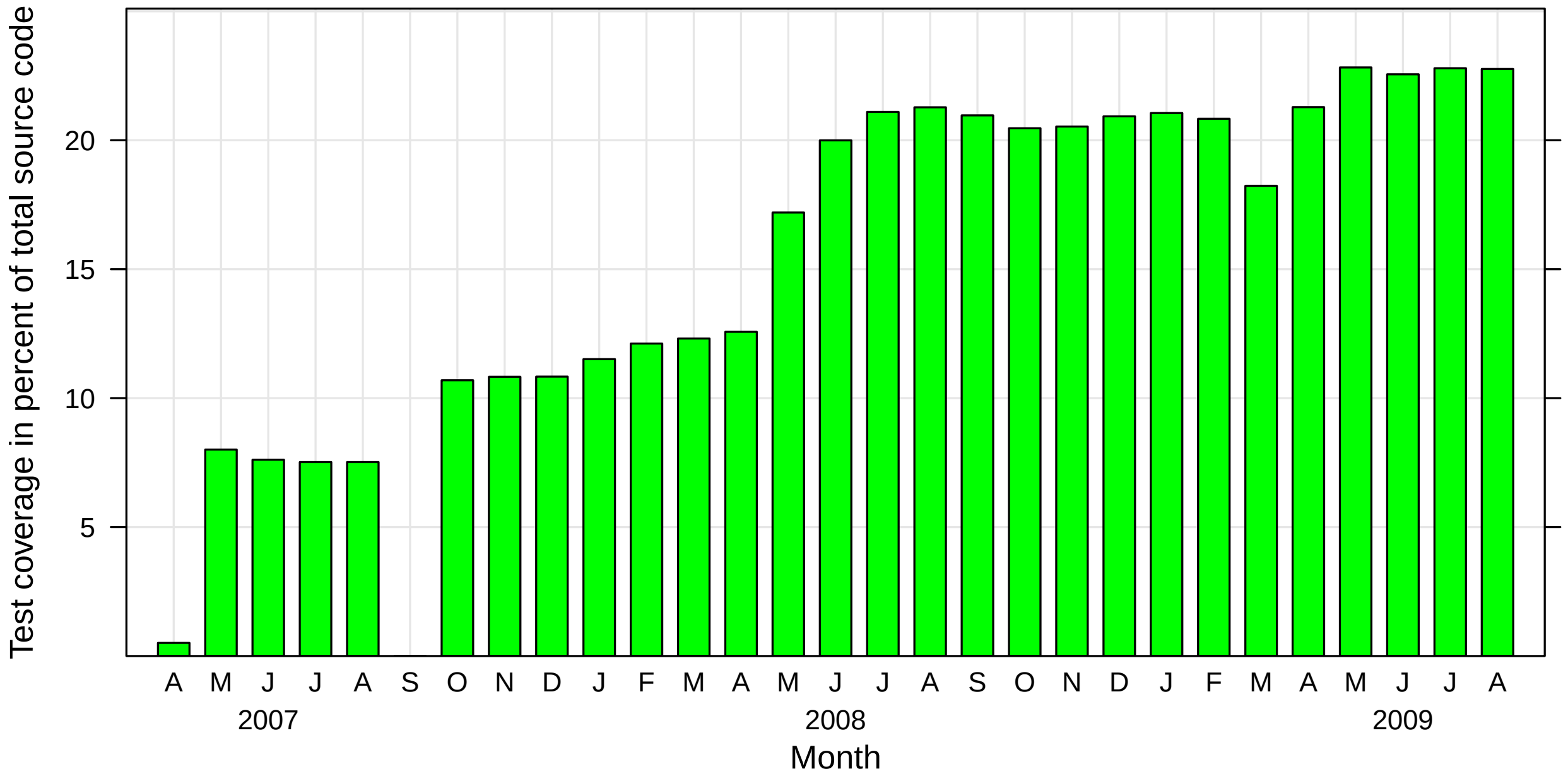

Figure 6. A time series display of the test coverage achieved over the course of the innovation introduction. Coverage is shown as the percentage of lines of code exercised by running the test suite over the total number of lines of code. We can take note that there are two very pronounced increases in coverage. One was induced by the innovator when introducing automated testing into the project in April to September of 2007, which resulted in the 10% margin being attained, and the second was in April to June 2008 when testing was hugely expanded to reach 20% coverage. After these jumps in coverage, there is a slow increase in coverage, reaching 23% in August 2009.

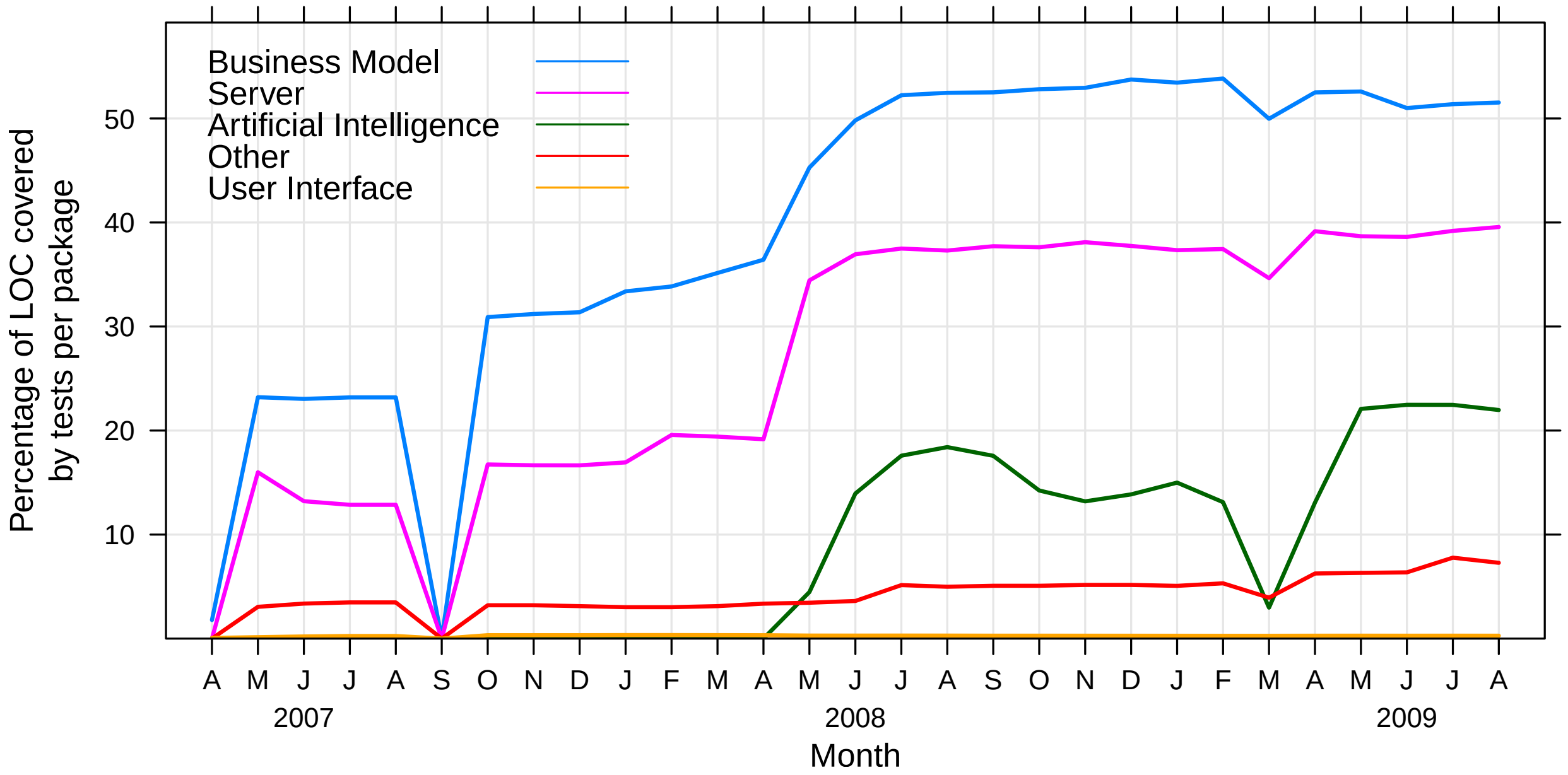

Figure 7. A time series display of the test coverage by major application module as a percentage of lines of code being exercised by running the test suite over the total number of lines of code. Noticeably, the application module most tested is the business model with around 50% which accounts for 7,400 lines of code being tested, with server code coming second at 40%. and the artificial intelligence routines at 22%. Also striking is the absence of any user interface testing (only 58 of 20,937 lines are tested in the latest check-out). The drop in coverage in March 2009 is induced by fifteen failing test cases (see Figure 8) and the missing results for September 2007 where the test suite was so broken that it did not compile.

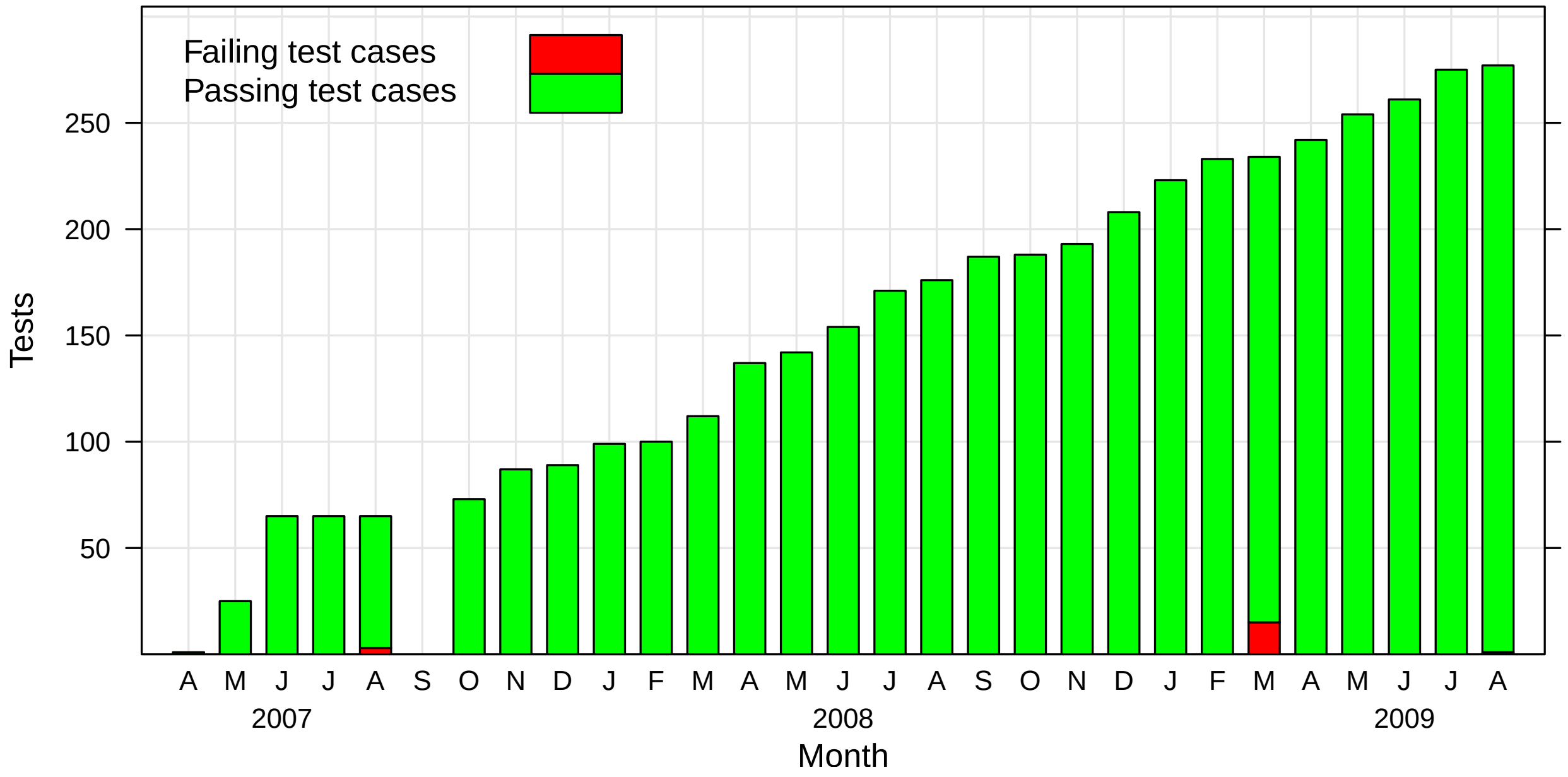

Figure 8. A time series display of the number of test cases in the FreeCol test suite distinguishing passing and failing tests on the first commit for each month. It can be seen that a.) the number of test cases is expanding on average by 9.9 test cases per month (median 9.0, standard deviation 8.9) and that b.) the source code is passing the test suite with very low numbers of failing tests. One notable exception to the latter is September 2004 when the whole test suite did no longer compile.

Instead, we want to (1) assess the magnitude of code-centric "hard" contributions vs. communication-based "soft" contributions and (2) abstract from the innovation of automated testing to innovations in general.

If we turn to the first aspect, we can note that if one is to distinguish between the project activities of communicating on the mailing-list ("soft" contributions) and committing to the project repository ("hard" contributions), then a slightly higher ratio between the number of discussion threads and the total commits (848 threads containing 3,163 e-mails over 3,676 commits, i.e. a ratio of 4.3 commits to 1 thread) and the number of threads relating to testing and the test affecting commits (71 threads containing 423 e-mails over 441 commits, i.e. 6.2 commits to 1 thread) can be found. These ratios are in line with results from literature [66] and help to put the communication aspect of testing into perspective: Open Source projects are focused on hard contributions (removing user requests, off-topic discussions and SPAM should bring the general ratio even more closely to the testing one as well) and the majority of commits are never discussed on the mailing-list, a phenomenon which has been called the *bias for action* of Open Source software development [78, pp.334f]. Thus one possible explanation for the relative success of testing in an Open Source project could derive from bridging hard and soft contributions via test cases which can be committed to the project repository while at the same time offering the possibility for communication.

As a second aspect and abstracting from testing as a particular innovation, we can deduce that *reinvention* — the usage of an innovation in an unexpected way [62] — can provide valuable benefits to the project not obvious to the innovator. Testing in this case had been targeted by the innovator at improving (1) the quality of the software produced in the Open Source project, and (2) its resistance against regression. Yet, it was used instinctively and in addition as a means of communicating complex expectations against the code or the behavior of others inside the project.

**Strategy 1.** *The innovator should keep an open, supportive mind about innovations being reinvented to gain the most tangible benefits for a project.*

As a second insight we found that testing varies largely by module, based on its technical complexity regarding testing. While the FreeCol business logic including the game objects attained more than 50% coverage, other areas such as the server module at 40% and the artificial intelligence module at 22% are less tested and UI testing is absent completely from FreeCol (see Figure 7). How to expand the coverage of underrepresented modules substantially is an open question for automated and unit testing in particular.

Finally, a failing of the whole test suite due to insufficient memory being allocated when running it — an event which occurred no less than five times over the two years — highlights the relationship between testing and specifying a software. Little memory was assigned primarily to the tests to prevent a memory leak from slipping into the software, but each failure due to insufficient memory also caused the question whether the failure was due to FreeCol having actually outgrown the memory limitations assigned to it [freecol:4276]. Developers needed to assess each time whether a failure is an acceptable consequence of FreeCol being expanded or a regression necessary to be rectified. This ambiguity regarding the specification of FreeCol then turned out to be difficult to resolve for the developers, one of whom remarked: "I don't remember why we decided to restrict memory during tests. I'm not sure whether it is a good idea, or not." [freecol:4272]. One conclusion for the innovator should be that the role of testing with regard to specifying the behavior of the software is explained in more detail to the developers so as to reduce the ambiguity if the specification can only be vague such as with limited memory.

## 4.2 Insights into innovation introduction

On introducing innovations we have found two main results in this study for projects comparable to FreeCol (see Section 5 on generalizability):

1. Open Source projects excel at incrementally expanding innovation usage over long time and maintaining an existing code base, yet does require assistance by an innovator or particularly skilled individual to achieve radical expansion with regard to an innovation.

2. When detaching from an Open Source project, the innovator should signal this to release ownership of responsibilities and code.

The first insight was deduced by analyzing the increase in coverage in the project (see Figure 6), which shows two notable expansions over the last two years. The first was the expansion of coverage from 0.5% to 10% by this author (as the innovator) when introducing automated testing to the project in 2007, and the second in April and May of 2008 when one developer expanded coverage from 13% to 20% by starting testing the artificial intelligence module (see Figure 7). Yet, beside these notable increases, which occurred over a total of only four months, coverage remained relatively stable over the two years. This is unlike the number of test cases which constantly increased with a remarkable rate of passing tests (see Figure 8). On the mailing-list a hint can be found that this is due to the difficulty of constructing scaffolding for new testing scenarios (see for instance the discussion following [freecol:4147] as to the problems of expanding client/server testing) and thus indirectly with unfamiliarity with testing in the project. This thus poses a question to our understanding of Open Source projects: If – as studies consistently show – learning ranks highly among Open Source developers' priorities for participation [25, 32, 34], then why is it that coverage expansion was conducted by just two project participants? Even worse, it seems that the author as an innovator and the one developer both brought existing knowledge about testing into the project and that project participants' affinity for testing and their knowledge about it expanded only very slowly.

**Conjecture 1.** *Knowledge about and affinity for innovations by individual developers is primarily gathered* outside *of Open Source projects.*

This conjecture can be strengthened by results from Hahsler who studied adoption and use of design pattern by Open Source developers. He found that for most projects only one developer — independently of project maturity and size up until very large projects — used patterns [27, p.121], which should strike us as strange if sharing of best practices and knowledge did occur substantially.[15] This argument then on the one hand can thus emphasize the importance of an Schumpeterian entrepreneur or innovator who pushes for radical changes:

**Conjecture 2.** *To expand the use and usefulness of an innovation radically, an innovator or highly skilled individual needs to be involved.*

On the other hand, if we put the innovator into a more strategic role as to acquire such highly skilled and knowledgeable project participants with regard to an innovation, then we can deduce two different options:

1.) Taking an optimistic view of the technical and intellectual skill of the developers participating in an OSS project, we might conclude for the innovator that additional training and support is necessary to counteract the lack of progress with regard to learning about a new innovation.

[15] Certainly to really confirm such a conjecture an in-depth study or interviews with the developers would be necessary [28].

**Strategy 2.** *The innovator needs to actively promote learning about an innovation to achieve levels of comprehension necessary for radical improvements using this innovation.*

This strategy has been used only very little by the innovator by providing a tutorial video about setting up Eclipse for testing and by writing a small tutorial document explaining how to write test cases and explaining some rationales for it. Yet beyond these actions by the innovator, no other attempts were made to systematically learn about the innovation. This poses an open question for Open Source research to which we can only suggest some initial ideas:

**Open Question 1.** *How can knowledge acquisition and sharing in Open Source projects be maximized?*

This open research question needs to be seen in addition to the results of existing research about information generated *inside* the project (see [55] for an innovation targeting increased information management capabilities in a project) such as information about tasks, social relationships and contextual factors [10]. It was found that Open Source projects are well positioned for such information because of their nature as Communities of Practice [41, 73, 70] which foster both re-experience and participatory learning [33, 32] and by use of information management infrastructures such as bug trackers, source code management systems or wikis [43]. Yet, we find that for innovation introductions and radical changes, the knowledge acquisition from the *outside* must be considered separately. This study did not aim to answer this question, but some general ideas come to mind if assuming the role of an at least somewhat proficient innovator and pondering how to increase knowledge about an innovation: First, the most obvious way and certainly the status quo in an Open Source project for the innovator to enhance knowledge is by explaining and tutoring the other project members via e-mail and references to external literature. Yet, presenting information in a static, non-interactive way via e-mail or the web has several drawbacks, such as for instance being more work for the author than verbal communication, equipped with less cues about which aspects are particular important, and with greater ambiguity about the success of transferring information to the recipients' side [10, p.220f]. To alleviate these problems, the innovator might want to strive for a more direct communication channel of which two particular kinds exist: (1) Real world meetings and conferences such as the *Free Software Developer European Meeting* (FOSDEM) or the KDE project's yearly aKademy[16] can provide a yearly get-together in which information about a new innovation can be distributed to all central project participants. (2) With the increasing maturity of collaboration awareness and screen sharing solutions, it becomes possible to conduct distributed pair programming (DPP) sessions for demonstration and teaching purposes (see for instance [3, 67, 30] for empirical results, [35, 51, 65, 15] for several tool solutions supporting DPP, and [77] for a comparison of tools) which can provide an efficient high-bandwidth synchronous communication channel between a knowledgeable project participant and those to be taught.

2.) Taking a more pessimistic view on the other hand, we might discard such a teaching approach when noting high turn around of participants in Open Source projects, which would foil any long term strategies to spread knowledge in OSS projects. Rather the innovator should focus on the acquisition of new developers already skilled with desired innovations from outside the project and focus on their inclusion into the project.

**Strategy 3.** *An innovator can strengthen innovation introductions by lowering entry barriers and acquiring highly skilled individuals.*

The second insight regarding innovation introduction arose from the circumstance and events surrounding the phasing-out of the innovator's involvement both in May and September 2007. As previously discussed, the first attempt of detaching from the project failed and the test suite as a consequence was unmaintained during a large-scale refactoring and thus soon "spectacularly broken", as one of the maintainers stated. Comparing this phase-out with the second much more successful one in September 2007, which resulted in the role of maintaining the tests being picked up by one of the maintainers, we find that the primary difference in behavior is one of signaling and ownership. When I — as the innovator — first detached from the role of managing test cases, I did neither consider ownership of the test code just being created nor communicating my withdrawal to the project an important priority. In fact, I did not consider myself the code owner of the testing code. Yet, just as Mockus et al. found in their case study of Apache and Mozilla, code ownership is achieved implicitly for code the developer is "known to have created or to have maintained consistently" [50, p.318]. While such code ownership "doesn't give them [the owner] any special rights over change control", it stipulates a barrier for other developers to engage with the code (see for instance [40] for a discussion on code ownership as an important part of the mental model of developers). Having not signaled the phase-out then prevented the other project participants from understanding that the test code should be perceived as under shared ownership. Shared ownership should be assumed because, unlike separate modules of the code being maintained, test code is deeply dependent upon the rest of the software as evident to it being broken completely due to the refactoring.

[16] For some analysis on participation in community events such as these, see [11, 6].

Only when the test suite decayed to be broken completely after the refactoring must it have become apparent that the suite was unmaintained. Abstracting from the case of introducing automated testing, this might have been a major factor in the success of the whole introduction: Running the test cases automatically provided a way to know whether the test suite was still being maintained by being signaled with a red bar of failing tests.

Thus, when phasing-out the innovator's engagement again after having fixed the test-suite in September 2007, a discussion (see [freecol:2182]) was sufficient to create this shared obligation for testing. When the innovator then disengaged, one of the maintainers picked up the role of maintaining the test cases successfully (see Figure 4), keeping the percentage of test affecting commits at around 10% of the total commits (see Figure 3) until another developer assumed a more active role in testing. We can conclude:

**Strategy 4.** *An innovator should explicitly signal his engagement and disengagement in the project to facilitate shared ownership of the innovation introduction. In particular this will be necessary if the innovation itself has no signaling mechanisms to show individual or general engagement or disengagement.*

When analyzing the contributions of developers to the testing effort, we find that besides myself as the innovator and the aforementioned maintainer there were two notable individuals who contributed extensively to testing. Interestingly, as their contribution increased and waned according to their time contribution to the project, the maintainer who had already picked up the testing effort from me seemed to adjust his own contribution to testing accordingly, which was best visible from January to May 2008, where the increasing contribution of one developer caused the maintainer to invest less into testing until the developer was inactive in June to August 2008[17] which caused the maintainer to resume testing activities (see Figure 4). As contributions of the other core and peripheral developers never exceeded five testing commits per month and never continued for an extended period of time, we can interpret that the project adopted a more flexible code ownership strategy. In this approach, the role of a "test master" exists who contributes heavily to testing and is pivotal to the expansion of test coverage and development of knowledge regarding testing. This role is not formally but rather implicitly assigned and acknowledged explicitly in the project only for instance when a core developer — stumped by a difficulty regarding testing — asked: "Any suggestions, particularly from the resident test expert [name of developer]?" [freecol:4446].

**Strategy 5.** *To maintain constant activity with regard to introducing an innovation in the face of fluctuating developer contributions, the project leadership and/or the innovator needs to be flexible to resume and cease their own contributions.*

[17] Due to vacation [freecol:2919].

## 5 Limitations and conclusion

In the last section, the limitations of this research regarding internal and external validity shall be discussed before summing up the results.

We first want to turn to external validity, i.e. the question whether the results of this study can be generalized to other settings, in particular whether the introduction of testing can be achieved in another project than FreeCol and with a different innovator. For other Open Source projects many aspects might be different to FreeCol, in particular (a) attitude toward testing and similar software process improvements, (b) a programming language other than Java might be used for which there is no well established automated testing framework such as JUnit, (c) the project might be differently sized, causing many other aspects of interaction to occur and have developers showing less interest in testing, and (d) the software produced by the project might have different characteristics regarding testing. For other innovators similarly (a) their level of knowledge about innovation introduction and (b) testing might be substantially different, as well as their (c) available time and (d) motivation to achieve the introduction successfully. We discuss these in turn.

- Different attitudes towards testing specifically and innovations in general might influence adoption and represents the most important threat to generalizability. If nobody would have found testing to be a worthwhile innovation the introduction would have failed with tests breaking as the source code continues to evolve. Looking at the case of FreeCol gives us one defenses against this threat: Contributing to testing seems to be an individual decision. This at least can be concluded by the large variability in testing commitment where some developers contribute while others do not. I then think it is reasonable to argue that (a) finding a developer interested in testing is correlated to project size and (b) that we did not see any indication in the case of FreeCol that the fraction should be lower in another project.
- If a different programming language than Java is used in a project in which testing is to be established, this may hurt external validity on several counts. For instance, one could argue that Java provides many features for modularizing a software into well-testable units, such as clearly separated interface definitions, package boundaries and visibility modifiers which might lack in other languages. Furthermore, one could argue that JUnit as a well-established automated testing framework created by even more well-known practitioners must make a Java project more likely to adopt testing. Arguing against this validity restriction, one could note that untyped programming languages such as Python, Ruby and Perl in particular have a tradition to replace the safety of a compiled language with strong test suites [17] and thus projects' using them should be even more inclined to use automated regression testing, and that all major programming language have comparable automated testing libraries such as CppUnit for C++, PerlUnit for Perl, PyUnit for Python, NUnit for C# or Test::Unit for Ruby [45, p.14].

- The size of the project as well as the characteristics of its participants might influence the results of the study to a degree which makes a transfer to other projects difficult. We think there are essentially three sub-types in this threat which need to be addressed: (1) The target project might be substantially smaller than FreeCol, (2) the project might be substantially larger than FreeCol, and (3) the participants in the project might have a negative attitude towards automated testing to start with. We do not believe that the first and second threat are fundamentally challenging the results of this study, because for projects sufficiently smaller than FreeCol (i.e. a maintainer and at most a handful of developers) the contribution of a single additional person such as an innovator will always have a much more marked influence on the project as in a smaller project. It is thus not likely that the contribution of the innovator is rejected and rather possible that the innovator will alone be able to achieve high levels of coverage within a short amount of time. Testing should then show its benefits in uncovering defects in the software and achieve adoption easier than with a medium-sized software such as FreeCol, where high levels of coverage are harder to attain. While achieving an initial introduction thus should be easier to attain in a smaller project, it should be more difficult to withdraw from the project because the number of developers available to assume the role of testing will be smaller. In a substantially larger project (i.e. around 20 core developers) we think it is reasonable to assume a higher level of software engineering knowledge and thus with a high probability either already an existing test suite or at least some experience with it. Both should increase the likelihood that an innovator willing to contribute test cases or expand on an existing test suite will be welcomed to do so. Conversely to a small project, we would argue that initial introduction might be harder but withdrawing would be easier. For the third threat, that of negatively minded developers, we must conclude that introduction will be more of a challenge and would primarily refer to existing research about how to deal with resistance to change in OSS projects [69]. Secondly, we should note that the introduction with FreeCol contained at least two aspects which mark it an introduction with some resistance: For one, the first attempt to withdraw failed and the innovator needed to expedite additional effort to achieve adoption, and second, only one of the maintainers supported the adoption actively while the other one had only a minor number of testing commits (see Figure 4). While an introduction might still be made much more difficult if no maintainer is championing the introduction, we think it reasonable that results should generalize.

- As a third threat to external validity one might note that a different kind of software being produced by the OSS project should lead to different adoption behavior. We agree it should, but we think that FreeCol as a desktop application including client/server network communication, artificial intelligence and complicated user interface should be at the one end regarding difficulty in testing. Other types of software such as for instance algorithmic and utility libraries, web frameworks [69] or programming languages should be far easier to test. Only in operating systems and systems libraries with their dependence on different configurations of hardware do we see a higher testing complexity than in desktop applications.

- Regarding validity concerns about the person of the innovator we must conclude that this author had probably more knowledge about how to introduce innovations than most innovators as described in Section 1. Yet, since he acted using the phase model as described in Section 3.1, it does not seem too difficult to replicate his actions in a different project. Rather, the insights presented in this study should already increase the likelihood for innovator success, for instance by appropriately signaling withdrawal from the project as described in Section 4.2. As for the other mentioned concerns of comparable motivation, testing knowledge and available time, we think them reasonable to assume for any innovator committed to establish testing in an OSS project. Time spent on introduction was deliberately kept at around 10-15 hours per week, and testing knowledge can be acquired before an introduction easily from existing technical literature. As for motivation, we can note that while this innovator was driven to achieve research results, the motivation by increased quality and probably a desire to participate in a project which one is dependent upon as assumed in the introduction should establish sufficient motivation.

Before moving on to the discussion about internal va-lidity, external validity of the results should be examined when abstracting from the concrete case of introducing the innovation of automated testing to the general set of innovations which could be introduced into an OSS project. To this end, testing can be categorized as (1) a knowledge intensive innovation with (2) medium up-front costs to establish benefit for a project, (3) high levels of trialability because of the orthogonality of testing technology with other infrastructure and pre-existing innovations, (4) high levels of independence regarding individual innovation decisions to adopt testing, (5) as requiring ongoing maintenance effort particular in the face of code refactorings, and (6) as transparently regarding the current state of the implementation of the innovation in the project as visible by the failure state of the test suite. Other interesting innovations, such as for instance a decentralized source code management system such as Git [29] or a quality assurance process of pre-commit peer reviews such as practiced by the Apache project [18], will likely differ in at least one of these aspects. For instance, for the introduction of a new source code management system, trialability is markedly reduced, as migrating the data from the existing repository to a novel one will either require duplicate effort to maintain both repositories or will lead to wasted efforts. For a process improvement such as peer review on the other hand, a different type of innovation decision is likely, because peer review in particular becomes useful if applied uniformly to all commits and thus should need a project-wide consensus[18]. Nevertheless it is hard to see why a result such as the importance of signaling departure or the strategic dichotomy between teaching existing project participants vs. recruiting new ones should not transfer to other innovation introductions.

Two major threats need to be discussed when turning now to internal validity: (1) This study relied exclusively on information which was publicly available in mailing-lists and the project source code management repository, possibly leading to conclusions deviating from the events as they actually occurred. (2) The long-term analysis performed on the repository and the mailing-list is coarsely grained by only performing monthly check-outs of the repository and by using keyword searches on the mailing-list.

The first threat arises from the methodology used, which is markedly in contrast to fieldwork and ethnographic studies conducted with companies (see for instance [42]). In this study we only regarded intermediates and process results, such as the mailing-list messages, source code commits and — to some limited extent — bug reports. One can argue against this being a real threat as the whole communication in the Open Source world is based on these forms of encoded information and no face-to-face communication exists. Thus, while we might not learn what one project participant actually did, we can nevertheless assume that most other project participants will have received a similarly restricted view on the participant's activities. If one would want to strengthen the research with regard to this threat, two major remedies can be offered: (a) Using a more collaboration-oriented research methodology such as action research based on a researcher client agreement [44, 68, 2, 12] should allow for a more detailed and comprehensive communication between the open source project participants and the researcher. Alternatively, research by West and O'Mahonycan be named as particular successful examples of using interviews with project members to derive interesting results about Open Source projects and their governance [56] and firm–community relations [75, 74]. (b) The researcher could use appropriate instrumentation in agreement with the project participants to capture more detailed information about their daily work processes. For instance using a tool such as Hackystat [37] or the ElectroCodeoGram [64], it should be possible to capture in detail each time a project participant executed test cases or made changes to them.A further threat to internal validity arises from the coarsely-grained data analysis. First, we only took monthly snapshots from the repository to graph the evolution of coverage and failure rate of the test suite over time[19]. A more fine-grained analysis could reveal notable stretches in-between those check-outs in which the test suite was broken, and should show in more detail how coverage was expanded by the project. Yet, we find it hard to see how this threat should attack the primary results drawn from the coverage and failure rate analysis, namely that coverage is steadily expanding and that the failure rate is actively controlled by the project members. Secondly, when searching the mailing-list for e-mails regarding automated testing, we employed a set of keywords[20] which might have missed important events on the mailing-list regarding testing which did not contain any of the keywords. While it is easy to guard against this threat by a more complete analysis of the messages sent to the mailing-list, we found that perusing over 3,100 e-mails a markedly less effective way than keyword searching.

When looking to future work it seems best to prioritize conducting a second case with another project that is different from FreeCol and then secondly use additional data sources such as in particular interview with project members to strengthen internal validity.

To conclude, this study has shown that the introduction of code-centric process innovation such as automated test-ing into an OSS project is feasible and can be successfully achieved. In particular, this study revealed that an innovator (1) from the outside of the project and (2) with a relatively short period of activity can trigger a beneficial innovation adoption curve based on a simple four-stage model of innovator activities.

[18]While the argument here is that peer review is likely to be adopted after a project-wide organizational innovation decision, Fogel notes an interesting introduction episode in the Subversion project in which one particular individual leading by example has achieved the introduction of peer review almost single-handedly [20, p.39ff].

[19]To be more precise: out of 3,776 commits during the observation period, only those 29 representing the first commit of each month were analyzed for coverage and test failures, which is less than 1% of commits.

[20]Keywords did include: junit, test*, regression, failure, automated

Regarding automated testing, this study has found a surprising number of episodes in which test cases were used for communicating bug reports and opinions about specification more precisely and as technical "hard" contributions. Secondly, the state of the practice regarding automated testing has been found to be lacking in modules such as the user interface, the artificial intelligence reasoner and client/server communication, which prevents further expansion in coverage and thus quality improvements.

For an innovator and innovation introduction in general, results agree with [55] in that reinvention can be a major source of benefit of an innovation, and that the innovator should thus seek or at least not prevent such from occurring. As to the radical expansion of innovation use in a project, two strategies are deducible from the events observed in the project FreeCol. These events highlight the importance of external participants for radical expansions, as both of the pushes in coverage were achieved by participants with knowledge about automated testing which precedes their involvement in FreeCol. This leads to the first strategy of reducing entry barriers for external developers as to leverage such pre-existing knowledge to the fullest. Yet, since inside the project the amount of coaching and teaching of automated testing specific knowledge did only occur minimally, we thus propose that an innovator or project leader could also attempt to employ such a teaching strategy to increase knowledge about testing and have coverage expanded as a result of the increased skill of existing project members.

As a third and last insight, signaling the departure of the innovator is important even for an innovation such as automated testing which has explicit signaling mechanisms such as test cases failing. Thus, when reducing effort or leaving the project, the innovator should inform the project and potentially even look for a successor for sustaining an innovation.

### 5.1 Acknowledgements

Dan Delorey provided the author with a list of all java projects on Sourceforge.net that had more than 5 active developers over the course of 2006. Many thanks also to Gesine Milde, Florian Thiel, Lutz Prechelt and the FreeCol maintainers and test masters who read a draft version of this paper.

## A Innovator Diary

- 2007-04-02 Checked-out project, subscribed to mailing-list and explored the application.

- 2007-04-03 Wrote a first test case to find a bug in the MapGenerator.

- 2007-04-04 Wrote a first e-mail to the mailing-list concerning test case for MapGenerator and accompanied it with a fix. This ended phase #1 (lurking) which only lasted 2 days thus and started phase #2 of actively contributing.

- 2007-04-05 The MapGenerator patch was included in FreeCol and I got a direct reply from Core Developer I of FreeCol. Began to structure a testing framework for FreeCol.

  Got a reply from the Core Developer I indicating that the project did not have a lot of exposure to automated testing so far. I volunteered to write a little introductory document regarding testing.

- 2007-04-06 Contacted the other maintainer (Maintainer I) for commit privileges after prompted by Core Developer I to do so.

- 2007-04-09 Received answer from Maintainer I. Commit privileges were granted to me, and I was introduced on the mailing-list.

- 2007-04-10 Wrote tests for pioneer work.

- 2007-04-11 Finished writing tests for pioneer work and committed them.

  Writing to the mailing-list about the deviations from the original game specification represented by these tests did elicit only a response from Core Developer I. While on technical level the discussion was successful in resolving the question of whether this deviation from the original FreeCol specification was intentional or not, from social perspective it would have been beneficial to achieve communication with the project more broadly.

- 2007-04-22 Running the test cases enabled me to catch a regression. In an e-mail to the mailing-list I mentioned the failure as a good example of how to write a test case to New Developer I. The next two weeks make up phase #3 of the innovation introduction and consisted of communicating and collaborating with others developers about testing.

- 2007-04-23 New Developer I used my suggestion and wrote two small test cases based on the code I had sent to him the previous day. The test case showed the unfamiliarity with testing in general for most of the people in the project. It helped to understand that there are several hurdles for the adoption of an innovation. In particular, his test cases returned the correct result and passed, but not for the reason he anticipated but rather because of incorrectly setting up the testing environment.

- 2007-04-23 Another New Developer II wrote a message to the list of being interested in helping out with the website, to which he got an reply one day later to write to the current webmaster and offer him his help.

- 2007-04-27 New Developer II made another join-attempt, this time based on an infrastructure suggestion (switching from the hosting platform SourceForge.net to Launchpad[21]) to which he got a short reply that his suggestion sounded nice but would be a lot of work, at which point the discussion stalled. This is indicative of a wrong join-strategy by newbies, who try to ask the members of the project for too much change.

- 2007-05-03 Used a bug report[22] to write twelve test cases of which 3 succeeded and 9 failed and posted it on the mailing-list. Directly got a reply by a lurker (New Developer III) who saw it as an opportunity to get involved. He said he had questions and I gave him my contact details (e-mail, IM).

- 2007-05-03 By now, a total of 57 tests have been written and it was deemed sufficient for self-sustaining growth. The next two weeks were decided to consistitute phase #4, phasing out the innovator's involvement.

- 2007-05-07 I produced a screencast of how to set up FreeCol for developing and regression testing in Eclipse.

- 2007-05-15 The build-scripts of the project were fixed to let those who do not use Eclipse for developing and

[21] Launchpad is a project hosting platform started by Canonical Inc. as part of their development on the Linux distribution Ubuntu. Homepage: `https://launchpad.net/`

[22] Bug #1616384 `https://sourceforge.net/tracker/?func=detail&atid=435578&aid=1616384&group_id=43225`

testing run the tests as well. This was intended to mark the end of my phase-out from FreeCol.

- 2007-05-16 The issues for which test cases were contributed on 2007-05-03 was fixed.
- 2007-05-22 Core Developer I was granted maintainer status and is now referred to as Maintainer II.
- 2007-05-30 Innovation introduction so far a failure. The number of test cases is stalled at 65 with no activity in the two weeks in which I did not participate. Without the engagement of the innovator no new test cases got written.
- 2007-08-28 I sent a literature suggestion to the list[23] which I thought had some relevance for a current design discussion and also declared my status as phased out. Maintainer II then informed me that this was a pity since the test cases were "spectacularly broken" since the refactoring of using a softcoded configuration. Figure 8 shows that since June the number of test cases had not increased anymore and that there was some decay in July already causing several failing tests at the beginning of August. The refactoring in August then caused both the highest number of commits per month in FreeCol ever (See Figure 2) and the whole test suite to break (as indicated by zero test cases for September 2007 in Figure 8).
- 2007-09-12 I asked the project how much they valued testing before agreeing to go about fixing the test cases broken by the restructuring.
- 2007-09-24 I repaired all test cases, but never indicated that I was intending to resume activities in the project. This thus can be seen as second attempt to phase out involvement.
- 2007-09-26 Helped Maintainer II to resolve an out-of-memory error, which occurred when running the test cases and was due to a memory leak in FreeCol.
- 2007-10-15 Wrote a test to be used for Test-Driven Development.
- 2008-01-09 Number of test cases surpasses 100, all of which were written by Maintainer II.
- 2008-01-12 A core developer who would become Test Master II later on (if the innovator is seen as Test Master I) contributed his first patch against the unit tests.
- 2008-05-18 A new developer who would later become Test Master III contributed his first commit for the unit tests.
- 2008-06-01 Number of test cases surpasses 150.
- 2008-10-14 New Developer III contacted me to ask about information about how to develop test cases.
- 2008-12-01 Number of test cases surpasses 200.
- 2009-05-01 Number of test cases surpasses 250.
- 2009-08-23 I started a full dump both of the FreeCol Subversion repository and requested a snapshot in *mbox* format of the mailing-list from Maintainer II.
- 2009-09-04 I finished the analysis of the Subversion logs and snapshot check-outs.

[23] Martin Fowler's "Dealing with Roles" [21]